\documentclass[preprint,nofootinbib,amsmath,amssymb]{revtex4}

\usepackage{graphicx}
\usepackage{multirow}

\begin{document}

\title{Confronting Tracker Field Quintessence with Data}

\author{Pao-Yu Wang}
\affiliation{Department of Physics, National Taiwan University, Taipei 10617, Taiwan, R.O.C.}

\author{Chien-Wen Chen}
\email{physcwchen@gmail.com} %
\affiliation{Leung Center for Cosmology and Particle Astrophysics
(LeCosPA), National Taiwan University, Taipei 10617, Taiwan, R.O.C.}

\author{Pisin Chen}
\affiliation{Kavli Institute for Particle Astrophysics and
Cosmology, SLAC National Accelerator Laboratory, Menlo Park, CA
94025, U.S.A.} \affiliation{LeCosPA, Department of Physics, and
Graduate Institute of Astrophysics,
National Taiwan University, Taipei 10617, Taiwan, R.O.C.} %

\begin{abstract}
We confront tracker field quintessence with observational data. The potentials considered in this paper include $V(\phi)\propto\phi^{-\alpha}$, $\exp(M_{p}/\phi)$, $\exp(M_{p}/\phi)-1$, $\exp(\beta M_{p}/\phi)$ and $\exp(\gamma M_{p}/\phi)-1$; while the data come from the latest SN Ia, CMB and BAO observations. Stringent parameter constraints are obtained. In comparison with the cosmological constant via information criteria, it is found that models with potentials $\exp(M_{p}/\phi)$, $\exp(M_{p}/\phi)-1$ and $\exp(\gamma M_{p}/\phi)-1$ are not supported by the current data.

\end{abstract}

\pacs{95.36.+x}

\maketitle

\section{INTORDUCTION\label{sec:introduction}}
Research on the cosmic accelerating expansion from both observational and theoretical sides has been proceeding intensely for the past dozen years (see~\cite{Frieman:2008sn,Caldwell:2009ix} for reviews). Within the framework of general relativity, the cosmic acceleration requires that the universe's current energy budget is dominated by a energy source, named dark energy, providing significant negative pressure density $p<-\rho/3$, where $\rho$ is its energy density. Combined constraints from different types of current observation~\cite{Komatsu:2010fb}$\textrm{--}$\cite{Vikhlinin:2008ym} render the best-fit dark energy equation of state, $w=p/\rho\sim-1$ with a few percent uncertainty, assuming w is constant. When dynamical $w$ is considered, its current value is still close to $-1$, with about $10\%$ uncertainty. This indicates that the cosmological constant remains the simplest valid realization of dark energy but there is still room for other possibilities.

Phenomenological studies on dark energy can in principle be categorized into two approaches. One is to reconstruct general properties of dark energy, the other is to constrain models on an individual basis. In the first approach, the evolution of equation of state $w(z)$, where $z$ is the redshift, for instance, can be reconstructed using either piecewise parametrization, continuous parametrization, or principal component analysis (see~\cite{Amanullah:2010vv},~\cite{Pan:2010rn} and~\cite{Clarkson:2010bm} for recent examples). The joint evolution of $w(z)$ and its time-derivative in units of the Hubble time, $w'=dw/dlna$, can also be reconstructed in comparison with theoretical boundaries for various dark energy classes~\cite{Chen:2009bca}. These studies render us general features of dark energy, yet the results may vary depending on the parametrization in use. Furthermore, the reconstruction technique has been developed to provide diagnostics~\cite{Sahni:2008xx} and consistency tests for dark energy models~\cite{Zunckel:2008ti}$\textrm{--}$\cite{Chen:2009xv}. While these tests can in principle be taken to falsify models, we should be aware of the possible bias carried by the the chosen parametrization~\cite{Chen:2009bca,Chen:2009xv}.

Orthogonal but complimentary to the reconstruction is the model-based approach, in which we directly constrain the parameter space of a dark energy model using observational data (see~\cite{Dutta:2006cf} for the case of pNGB quintessence, for example). As the precision of observation advances, this approach is becoming effective. Whereas robust and stringent constraints on model parameters can be obtained, a model's validity and its relative merit to other contending models can be evaluated via the goodness of fit (GoF) and the information criteria.

In this paper, we take the model-based approach to confront the tracker field quintessence model~\cite{Zlatev:1999tr,Steinhardt:1999nw} with observational data. In the quintessence scenario, the late time cosmic acceleration is driven by a scaler field witch slowly rolls down its potential. As a class of quintessence model, the tracker field has an attractive feature that there exists a common solution to the equation of motion, extremely insensitive to initial conditions. This feature can address the cosmic coincidence problem, that is, the ratio of the dark energy density to the matter density must be set to a specific, infinitesimal value in the early universe in order to be of the order of one today. The data we use come from the latest Type Ia supernova (SN Ia) compilation, the cosmic microwave background (CMB) observation, and the observation of the baryon acoustic oscillations (BAO). Besides constraining the model parameters, we assess the GoF and the model strength in comparison with the cosmological constant model.

\section{TRACKER FIELD QUINTESSENCE}

\subsection{Quintessence formalism}
In the quintessence scenario~\cite{Caldwell:1998ii}, the late time cosmic acceleration is driven by a dynamical scalar field $\phi$ slowly evolving in the potential $V(\phi)$. In a flat Friedmann-Robertson-Walker (FRW) universe, the evolution of
the scalar field is governed by its equation of motion
\begin{equation} \ddot{\phi}+ 3H\dot{\phi}+\frac{dV}{d\phi} = 0, \label{one}\end{equation}
where the dots denote derivatives with respect to time, $H$ is the Hubble expansion rate $\dot{a}/a$ (a is the scale factor) given by the Friedmann equation
\begin{eqnarray} H^2(z)&=&\frac{8\pi G}{3}\left[\rho_r(z)+\rho_m(z)+\rho_\phi(z)\right]\nonumber \\
&=&H_0^2 \left[ \Omega_{r} (1+z)^4+\Omega_{m} (1+z)^3 +  \Omega_{\phi} \exp
\left( 3 \int_{0}^{z} \left[ 1 + w_{\phi}(z') \right]
\frac{dz'}{1+z'} \right) \right],
\label{two}\end{eqnarray}
where $\rho_r(z)$ is the radiation energy density, $\rho_m(z)$ is the matter energy density, $\rho_\phi(z)$ is the scalar field energy density, $H_0$ is the Hubble constant and $w_{\phi}(z)$ is the equation of state of the scalar field. The total fractional energy density today $\Omega_{total}=\Omega_{r}+\Omega_{m}+\Omega_{\phi}$ is equal to one in a flat universe. The energy density and pressure density of the scalar field are
\begin{equation} \rho_{\phi} =
\frac{1}{2}\dot{\phi}^2 + V(\phi) \label{three},\end{equation}
\begin{equation} p_{\phi} =
\frac{1}{2}\dot{\phi}^2 - V(\phi)\label{four}.\end{equation}
The equation of state $w_{\phi}(z)=p_{\phi}/\rho_{\phi}$ changes with time and becomes negative when the potential is dominant. In the limit when $\dot{\phi}^2\ll V(\phi)$, the scalar filed has $w_{\phi}(z)\sim-1$.

\subsection{Tracker fields}
Tracker fields are a class of quintessence that address the coincidence problem. In these models, a wide range of initial conditions in the early universe evolve toward a common solution, called tracking solution, giving the same late time evolution of $\phi$, $w_{\phi}$ and $\rho_\phi$, and allowing the scalar field to drive the cosmic acceleration. The central theorem in~\cite{Steinhardt:1999nw} states that tracking behavior with $w_\phi < w_B$, where $w_B$ is the equation of state of the dominant background component, occurs for any
potential in which  $\Gamma \equiv V''V/(V')^2 >1 $ (the primes denote derivatives with respect to $\phi$) and
is nearly constant $|\Gamma'/\Gamma(V'/V)|\ll 1$ over the range of plausible initial conditions. The range of plausible initial conditions extends from $\rho_\phi$ equal to the initial radiation energy density $\rho_r$ in the early universe down to $\rho_\phi$ equal to the current matter energy density $\rho_m$. This constraint is necessary for the scalar field to converge to the tracking solution before the present time. The feature $w_\phi< w_B$ means that $\rho_{\phi}$ decreases more slowly then the background energy density. Eventually, at late time the scalar field density overtakes the matter density and becomes the dominant component.

\begin{figure}[ph]
\includegraphics[width=1\linewidth]{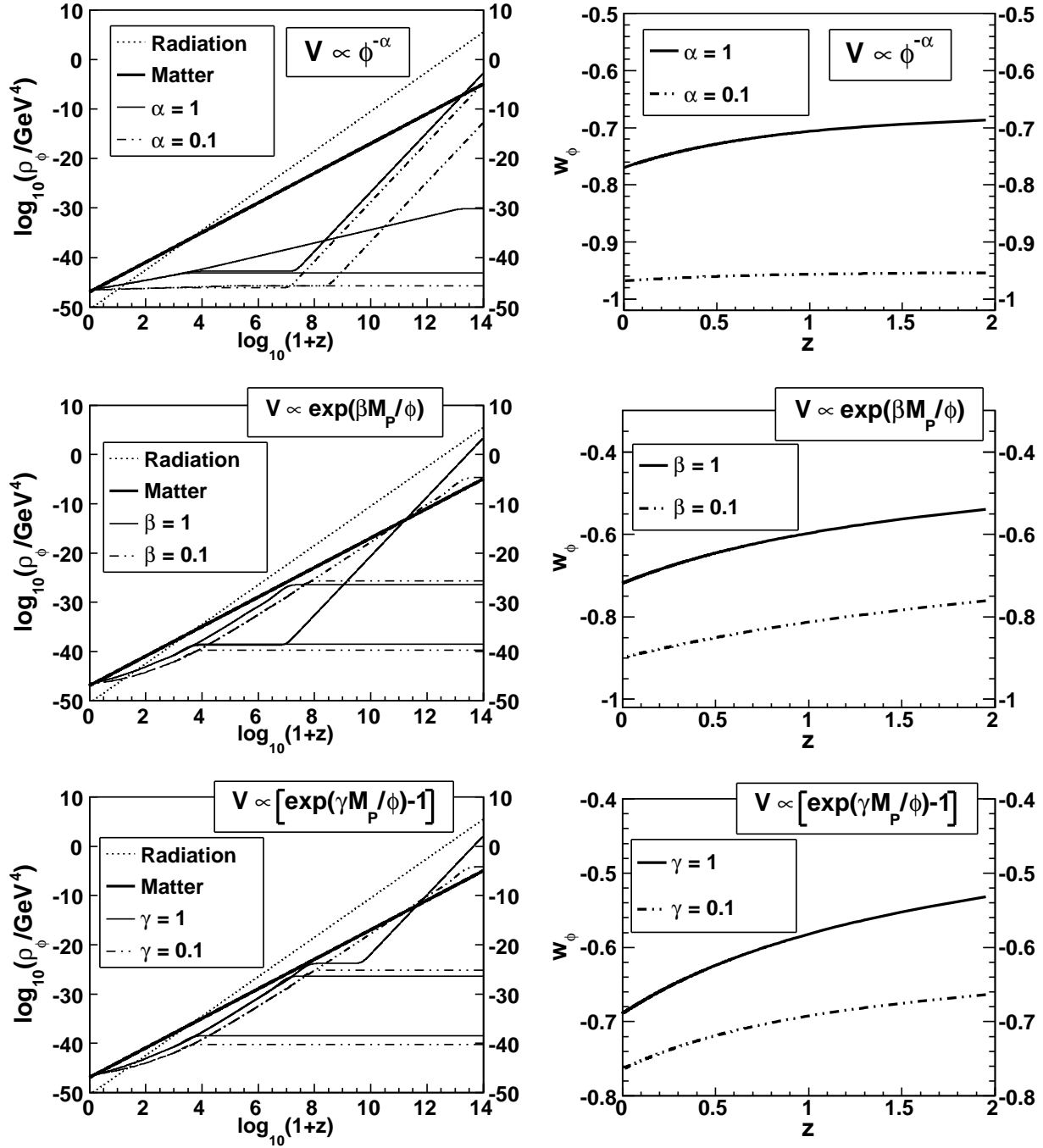}
\caption{An illustration of the tracking behavior and late time evolution of $w_{\phi}(z)$ of the tracking solution. The models include $V(\phi)\propto\phi^{-\alpha}$, $\exp(\beta M_{p}/\phi)$ and $\exp(\gamma M_{p}/\phi)-1$. $\Omega_m$ is set to be $0.27$. $\rho_{\phi}$ starting from a wide dynamical range in the early universe all converge to the same tracking solution at late time. For all the three models, $w_{\phi}(z)$ decreases in time at late time, the smaller the model parameter is the more negative value $w_{\phi}(z)$ can reach.
\protect\label{fig1}}
\end{figure}

A class of potentials that satisfies the conditions that $\Gamma>1$ and $\Gamma$ is nearly constant includes the inverse power-law (IPL) potential $V(\phi)\propto\phi^{-\alpha}$, and combinations of IPL terms, for example $V(\phi)\propto\exp(M_{p}/\phi)$, where $M_{p}$ is the Planck mass. Some of these potentials are motivated by particle physics models with dynamical symmetry breaking or nonperturbative
effects~\cite{Binetruy:1998rz}$\textrm{--}$\cite{Affleck}. The illustration in Fig.~\ref{fig1} exemplifies the tracking behavior and the late time tracking solution of these models. Details of the tracker field property can be found in~\cite{Zlatev:1999tr,Steinhardt:1999nw,Steinhardt:2005qf}. Analytical solution to the IPL model has been studied in~\cite{Watson:2003kk,Chiba:2009gg}.

In this paper, we analyze the models exemplified in~\cite{Steinhardt:1999nw}, $V(\phi)\propto\phi^{-\alpha}$, $V(\phi)\propto \exp(M_{p}/\phi)$, and $V(\phi)\propto \exp(M_{p}/\phi)-1$. We also analyze the generalization of the last two, $V(\phi)\propto \exp(\beta M_{p}/\phi)$ and $V(\phi)\propto \exp(\gamma M_{p}/\phi)-1$. All $\alpha$, $\beta$ and $\gamma$ are positive constant. Both $V(\phi)\propto \exp(M_{p}/\phi)$ and $V(\phi)\propto \exp(M_{p}/\phi)-1$ are distinct from the cosmological constant without extra parameters being introduced. $V(\phi)\propto\phi^{-\alpha}$ and $V(\phi)\propto \exp(\beta M_{p}/\phi)$ behave like a cosmological constant as $\alpha$ and $\beta$ approach zero, respectively. $V(\phi)\propto \exp(\gamma M_{p}/\phi)-1$ does not have a limit as the cosmological constant.

\section{Data}

We use observational data from SN Ia, CMB and BAO as described below.

\subsection{Type Ia supernovae}

We use the latest SNe Ia dataset, Union2 compilations, released by Supernova Cosmology Project which contains 557 SNe Ia in the the redshift range $0.02<z<1.5$~\cite{Amanullah:2010vv}. This compilation includes supernova data from~\cite{Hamuy:1996su}\textrm{--}\cite{Kessler:2009ys}. The dataset provides distance modulus which contains information of luminosity distance that can be used to constrain dark energy.

The distance modulus is defined as following:
\begin{equation}
\mu_{th}(z)=5log_{10}\left(\frac{d_L(z)}{Mpc}\right)+25=5log_{10}\left(D_L(z)\right)+\mu_0,
\end{equation}
where $D_L(z)=H_0 d_L(z)$ is Hubble-free luminosity distance. We marginalize $\chi^2_{SNIa}$ over the nuisance parameter $\mu_0$ by minimizing it with respect to $\mu_0$. The marginalized $\chi^2_{SNIa}$ is~\cite{DiPietro:2002cz}\textrm{--}\cite{Wei:2009ry}
\begin{equation}
\tilde{\chi}^2_{SNIa}=A-\frac{B^2}{C},
\end{equation}
where
\begin{eqnarray}
A=\sum_{ij}\left(5log_{10}\left[D_L(z_i,par)\right]-\mu_{obs}(z_i)\right)C^{-1}_{ij}\left(5log_{10}\left[D_L(z_j,par)\right]-\mu_{obs}(z_j)\right),
\end{eqnarray}
\begin{eqnarray}
B=\sum_{ij}\left(5log_{10}\left[D_L(z_i,par)\right]-\mu_{obs}(z_i)\right)C^{-1}_{ij},
\end{eqnarray}
\begin{eqnarray}
C=\sum_{ij}C^{-1}_{ij}.
\end{eqnarray}

\subsection{Cosmic microwave background}

The seven-year WMAP results provide "distance prior" that can be used to constrain dark energy~\cite{Komatsu:2010fb}. Distance prior includes CMB shift parameter $R=1.725\pm0.018$ given by
\begin{equation}
R=\sqrt{\Omega_m H_0^2}\left(1+z_*\right)D_A\left(z_*\right),
\end{equation}
and "acoustic scale" $l_A=302.09\pm0.76$ given by
\begin{equation}
l_A= \left(1+z_*\right)\frac{\pi D_A\left(z_*\right)}{r_s\left(z_*\right)},
\end{equation}
where $z_*$ is the redshift of decoupling, $D_A$ is the angular diameter distance, and $r_s$ is the comoving sound horizon. We use the fitting formula proposed by Hu and Sugiyama~\cite{Hu:1995en}:
\begin{equation}
z_*=1048\left[1+0.00124(\Omega_b h^2)^{-0.738}\right]\left[1+g_1(\Omega_m h^2)^{g_2}\right],
\end{equation}
\begin{eqnarray}
g_1=\frac{0.0783(\Omega_b h^2)^{-0.238}}{1+39.5(\Omega_b h^2)^{0.763}},
\end{eqnarray}
\begin{eqnarray}
g_2=\frac{0.560}{1+21.1(\Omega_b h^2)^{1.81}}.
\end{eqnarray}
The comoving sound horizon is
\begin{equation}
r_s(z)=\frac{1}{\sqrt{3}}\int^{1/(1+z)}_{0}\frac{da}{a^2H(a)\sqrt{1+(3\Omega_b/4\Omega_{\gamma})a}},
\end{equation}\\
where $\Omega_b$ is baryon density and $\Omega_{\gamma}$ is photon density.\\
We construct $\chi^2_{CMB}=\sum_{ij}(x_i-x^{Obs}_i)(C^{-1}_{ij})(x_j-x^{Obs}_j)$, where $C^{-1}_{ij}$ is the inverse covariance matrix given in~\cite{Komatsu:2010fb}, and $x_i=(l_A, R, z_*)$.

\subsection{Baryon acoustic oscillations}

We use BAO data from the joint analysis of Two Degree Field Galaxy Redshift Survey (2dFGRS) data~\cite{Cole:2005sx} and Sloan Digital Sky Survey (SDSS) Data Release 7 which provides two distance measures, $d_{0.35}=r_s(z_d)/D_V(0.35)=0.1097\pm0.0036$ and $d_{0.2}=r_s(z_d)/D_V(0.2)=0.1905\pm0.0061$~\cite{Percival:2009xn}, where $r_s(z_d)$ is the acoustic sound horizon at the drag epoch, $D_V= \left[(1+z)^2D^2_A(z)/H(z)\right]^{1/3}$. Fitting formula for $z_d$ is defined by Eisenstein \& Hu~\cite{Eisenstein:1997ik}. The $\chi^2_{BAO_1}$ is $\sum_{ij}(d_i-d^{obs}_i)(C^{-1}_{ij})(d_j-d^{obs}_j)$, where $d_i=(d_{0.2}, d_{0.35})$,
\begin{equation}
C^{-1}=
\left(
\begin{array}{cc}
30124 & -17227 \\
-17227 & 86977 \end{array}
\right).
\end{equation}
The fitting formula for $z_d$ has this form:\\
\begin{equation}
z_d=\frac{1291(\Omega_mh^2)^{0.251}}{1+0.659(\Omega_mh^2)^{0.828}}\left[1+b_1(\Omega_bh^2)^{b_2}\right],
\end{equation}
\begin{eqnarray}
b_1=0.313(\Omega_mh^2)^{-0.419}\left[1+0.607(\Omega_mh^2)^{0.674}\right],
\end{eqnarray}
\begin{eqnarray}
b_2=0.238(\Omega_m h^2)^{0.223}.
\end{eqnarray}

We also include BAO result from WiggleZ Dark Energy Survey~\cite{Blake:2011wn}, which gives $A(0.6)=0.452\pm0.018$. A(z) is given by
\begin{equation}
A(z)=\frac{D_V(z)\sqrt{\Omega_m H_0^2}}{z}.
\end{equation}
 The $\chi^2_{BAO_2}=[A(z)-0.452]^2/0.018^2$. Therefore, $\chi^2_{BAO}= \chi^2_{BAO_1}+ \chi^2_{BAO_2}$.
\subsection{Prior}
 For the radiation, we fix $\Omega_{\gamma}=2.469\times10^{-5}/h^2$, and the radiation energy density $\Omega_r=\Omega_{\gamma}(1+0.2271N_{eff})$, where $N_{eff}$ is the effective number of neutrino species and is taken to be $3.04$~\cite{Komatsu:2008hk}. We further impose the prior of $H_0=73.8\pm2.4~km s^{-1} Mpc^{-1}$ from~\cite{Riess:2011yx}. The total chi-square $\chi^2_{total}=\tilde\chi^2_{SNIa}+\chi^2_{CMB}+\chi^2_{BAO} + \chi^2_{H_0}$ is marginalized over the nuisance parameters $\Omega_bh^2$ and the reduced hubble constant $h$, by minimizing $\chi^2_{total}$ with respect to $\Omega_bh^2$ and $h$~\cite{Komatsu:2008hk}.

\section{Observational constraints on tracker field models}
The models we analyze include $V(\phi)=M^{4+\alpha}\phi^{-\alpha}$, $V(\phi)=M^{4}\exp(M_{p}/\phi)$, $V(\phi)=M^{4}[\exp(M_{p}/\phi)-1]$, $V(\phi)=M^{4}\exp(\beta M_{p}/\phi)$, and $V(\phi)=M^{4}[\exp(\gamma M_{p}/\phi)-1]$. All $\alpha$, $\beta$ and $\gamma$ are positive constant. The mass $M$ is determined by requiring that the total fractional energy $\Omega_{total}$ equals to $1$ in a flat universe. The initial conditions of $\phi$ and $\dot{\phi}$ are arbitrarily chosen in the range ensuring that the scalar field joins the tracking solution before the last scattering. We solve Eq.~\ref{one} and Eq.~\ref{two} numerically in order to calculate the chi-square for each point in the parameter space.

\begin{figure}[ph]
\includegraphics[width=1\linewidth]{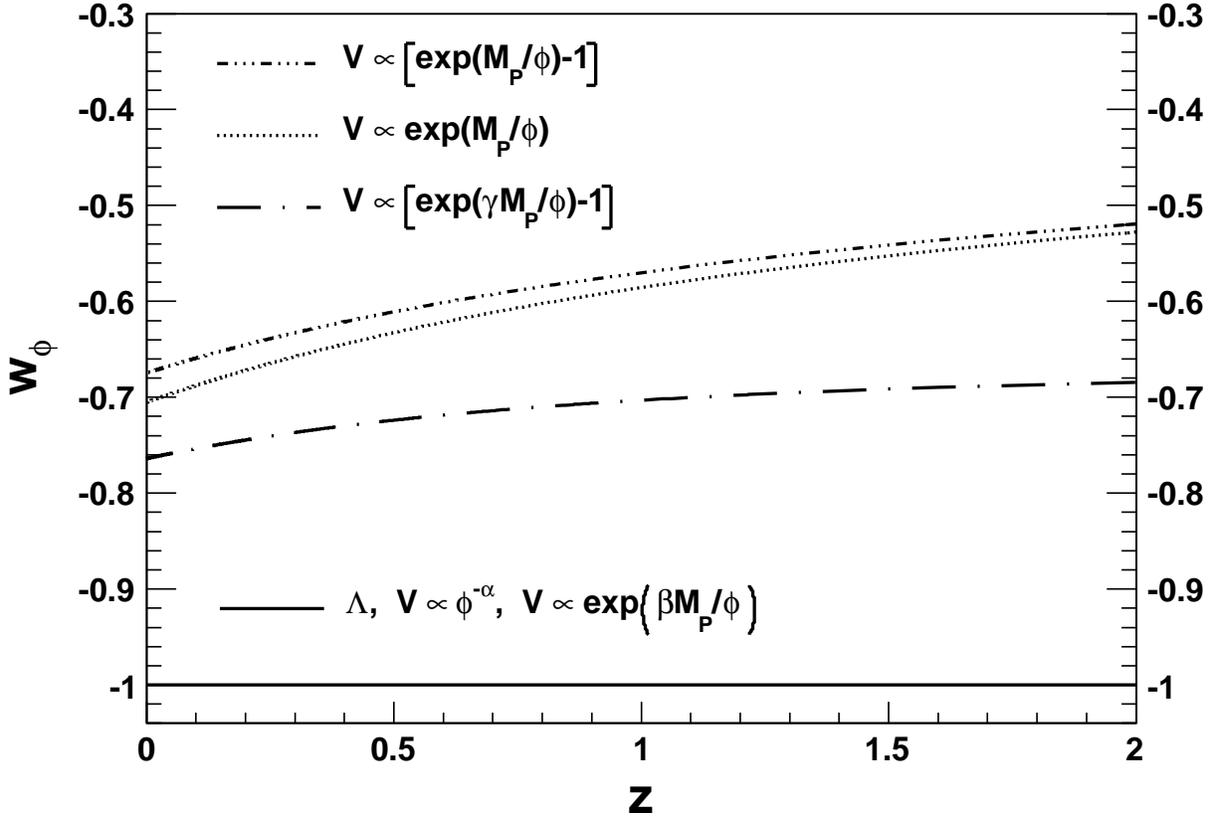}
\caption{The late time evolution of $w_{\phi}(z)$ corresponding to the best-fit parameters. Models with $V(\phi)\propto \phi^{-\alpha}$ and $V(\phi)\propto \exp(\beta M_{p}/\phi)$ have their best-fit acting as the cosmological constant ($\Lambda$). Models with $V(\phi)\propto \exp(M_{p}/\phi)$, $V(\phi)\propto [\exp(M_{p}/\phi)-1]$ and $V(\phi)\propto [\exp(\gamma M_{p}/\phi)-1]$ have their $w_{\phi}(z)$ of the best-fit staying away from $-1$.
\protect\label{wbest}}
\end{figure}

\begin{figure}[ph]
\includegraphics[width=1\linewidth]{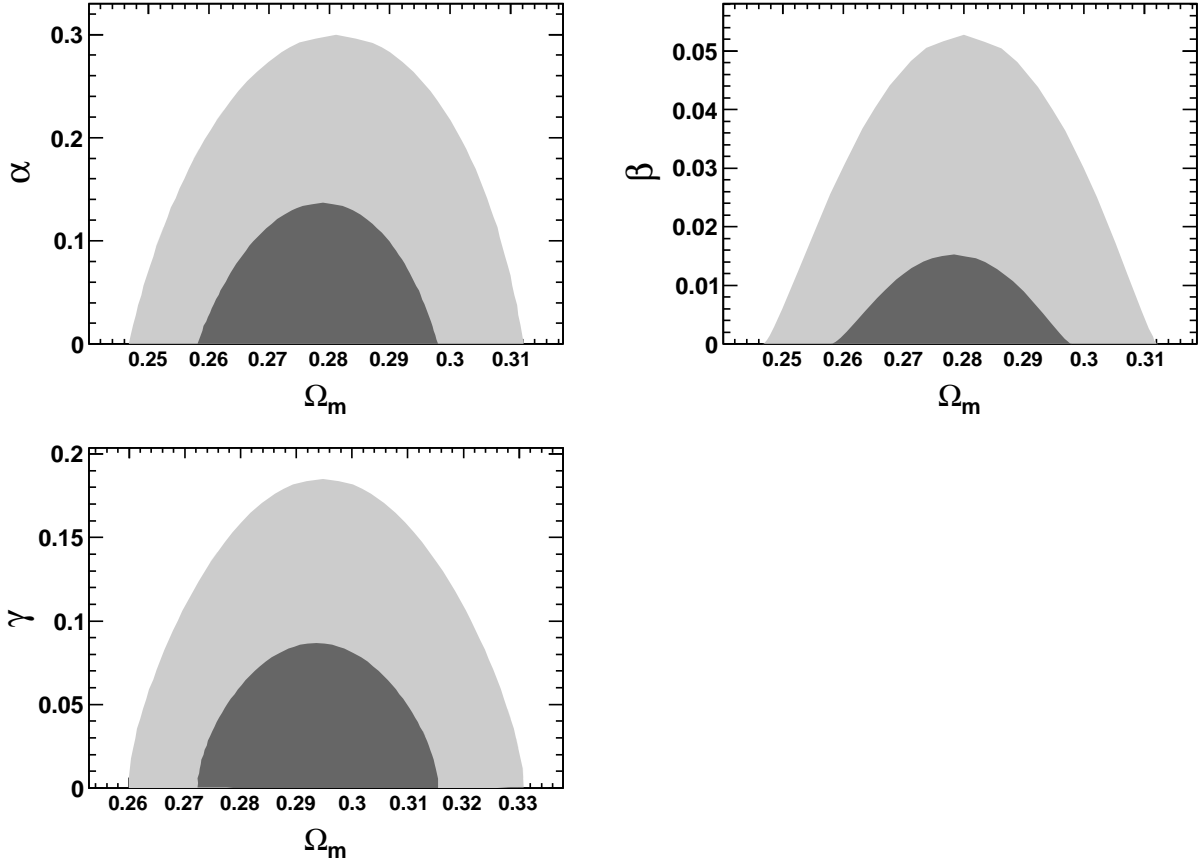}
\caption{Joint constraints on ($\Omega_m, \alpha$), ($\Omega_m, \beta$), and ($\Omega_m, \gamma$). The dark gray and the light gray regions correspond to the $68.3\%$ and $95.4\%$ confidence regions, respectively.
\protect\label{joint}}
\end{figure}

The resulting best-fit parameters for these five models are listed in Table~\ref{table:results}. The late time evolution of $w_{\phi}(z)$ corresponding to the best-fit parameters are plotted in Fig.~\ref{wbest}. The joint constraints on ($\Omega_m, \alpha$), ($\Omega_m, \beta$), ($\Omega_m, \gamma$) are shown in Fig.~\ref{joint}.

\begin{table}[ht]
\caption{Fitting Results}
\centering
\begin{tabular}{c|c c c c}
\hline\hline
Model & Best-fit parameters & GoF & $\Delta$BIC & $\Delta$AIC \\
\hline
Cosmological Constant & $\Omega_m=0.277^{+0.013}_{-0.013}$ & $76.9\%$ & 0 & 0 \\
\hline
\multirow{2}{*}{$V\propto\phi^{-\alpha}$} & $\Omega_m=0.277^{+0.013}_{-0.013}$ &\multirow{2}{*}{$76.0\%$} & \multirow{2}{*}{6.3} & \multirow{2}{*}{2.0} \\
& $\alpha=0^{+0.070}$ & & & \\
\hline
\multirow{2}{*}{$V\propto \exp\left(\beta M_P/\phi\right)$} & $\Omega_m=0.277^{+0.013}_{-0.013}$ &\multirow{2}{*}{$76.0\%$} & \multirow{2}{*}{6.3} & \multirow{2}{*}{2.0} \\
& $\beta=0^{+0.0051}$ & & & \\
\hline
\multirow{2}{*}{$V\propto \left[\exp\left(\gamma M_P/\phi\right)-1\right]$} & $\Omega_m=0.293^{+0.015}_{-0.014}$ &\multirow{2}{*}{$41.0\%$} & \multirow{2}{*}{37.2} & \multirow{2}{*}{32.9} \\
& $\gamma\to0^{+0.049}$~\footnote{the best-fit $\gamma$ arbitrarily approaches $0$} & & & \\
\hline
$V\propto \exp\left(M_P/\phi\right)$ & $\Omega_m=0.301^{+0.015}_{-0.015}$ &$16.5\%$ & 57.5 & 57.5 \\
\hline
$V\propto \left[\exp\left(M_P/\phi\right)-1\right]$ & $\Omega_m=0.307^{+0.016}_{-0.015}$ &$11.7\%$ & 65.1 & 65.1 \\
\hline\hline

\end{tabular}

\label{table:results}
\end{table}

We further evaluate the GoF\footnote{Defined as GoF=$\Gamma(\nu/2,\chi^2/2)/\Gamma(\nu/2)$, where $\Gamma(\nu/2,\chi^2/2)$ is the upper incomplete gamma function and $\nu$ is the degrees of freedom.} for each model (see Table~\ref{table:results}). The GoF gives the probability of obtaining data that are a worse fit to the model based on $\chi^2$ statistics, assuming that the model is correct. It can test the validity of a particular model. To assess the relative model strength, especially in comparison with the cosmological constant, we use the information criteria (IC). The IC are model selecting statistics encoding the tension between quality of fit and model complexity. They favor models that give a good fit with fewer parameters. The use of IC in the context of cosmological observation has been examined in~\cite{Liddle:2004nh}. In this paper we evaluate the Bayesian information criterion (BIC)~\cite{Schwarz} and the Akaike information criterion (AIC)~\cite{Akaike} for each model.

The BIC is defined as $\textrm{BIC}=-2\ln \mathcal{L}_{max} + k \ln N$, where $\mathcal{L}_{max}$ is the maximum likelihood, which is equivalent to the minimum $\chi^2$ for gaussian errors, $k$ is the number of parameters, and $N$ is the number of data points used in the fit. It comes from
approximating the evidence ratios of models, known as the Bayes factor. A better model has a lower BIC. The AIC is defined as $\textrm{AIC}=-2\ln \mathcal{L}_{max} + 2k$. The AIC is derived by an approximate minimization of the Kullback--Leibler information entropy, which measures the difference between the true data distribution and the model distribution. A better model has a lower AIC. The BIC gives stiffer penalty for extra parameters for the size of data $\ln N > 2$. The differences in BIC ($\Delta$BIC) and AIC ($\Delta$AIC) between each tracker field model and the cosmological constant are listed in Table~\ref{table:results}.

\section{CONCLUSION}

We have examined tracker field models with the potentials $V(\phi)\propto\phi^{-\alpha}$, $V(\phi)\propto \exp(M_{p}/\phi)$, $V(\phi)\propto \exp(M_{p}/\phi)-1$, $V(\phi)\propto \exp(\beta M_{p}/\phi)$ and $V(\phi)\propto \exp(\gamma M_{p}/\phi)-1$, based on current observational data. It is shown that the resulting parameter constraints are stringent (see Table~\ref{table:results} and Fig.~\ref{joint}). Best-fit of the two models $V(\phi)\propto\phi^{-\alpha}$ and $V(\phi)\propto \exp(\beta M_{p}/\phi)$ are equivalent to the cosmological constant (see Table~\ref{table:results} and Fig.~\ref{wbest}). The best-fit of the other three models that do not have limits as the cosmological constant render late-time equation of state staying away from $-1$ ($w_{\phi}>-0.8$). The larger values of $w_\phi$ come from the intrinsic limits of these models.

The poor GoF ($\textrm{GoF} < 17\%$) of models $V(\phi)\propto \exp(M_{p}/\phi)$, $V(\phi)\propto \exp(M_{p}/\phi)-1$ indicates these two models are less valid. The rank of model strength is the same assessed either by BIC or AIC. In comparison with the cosmological constant, the three models $V(\phi)\propto \exp(M_{p}/\phi)$, $V(\phi)\propto \exp(M_{p}/\phi)-1$, and $V(\phi)\propto \exp(\gamma M_{p}/\phi)-1$ have $\Delta \textrm{BIC}\gg6$, while $\Delta \textrm{BIC}>6$ is consider a strong evidence against the model~\cite{Liddle:2004nh}. This shows that the worthiness of considering these three models, in the presence of the cosmological constant, is not supported by the current observational data.

The result that both $V(\phi)\propto\phi^{-\alpha}$ and $V(\phi)\propto \exp(\beta M_{p}/\phi)$ have the best-fit as the cosmological constant suggests that other dark energy models which do not have the boundary $w>-1$ might render better fits to the data. These include the phantom models~\cite{Caldwell:1999ew} with $w<-1$, the quintom models (see~\cite{Cai:2009zp} for a review) with $w$ crossing $-1$ and the K-essence models (see~\cite{Malquarti:2003nn} for a review). The observational constraints on these models and their model strength should be further studied.

The next generation cosmological probes are expected to constrain $w$ about ten times better~\cite{Albrecht:2006um}. More stringent constraints on individual models are also expected to be obtained in the future (see~\cite{Barnard:2007ta}$\textrm{--}$\cite{Yashar:2008ju} for the case of quintessence models). In the ongoing pursuit of revealing the nature of dark energy, the reconstruction of the general features and the model based approach should be complimentary to each other. While testing the cosmological constant by examining if $w=-1$ and if $w$ has dynamical behavior, we should also take the model based approach to see if there is other model worth considering.

\begin{acknowledgments}
This work is supported by the Taiwan National Science Council under Project No. NSC
97-2112-M-002-026-MY3 and by US Department of Energy under Contract
No. DE-AC03-76SF00515. We thank Leung Center for Cosmology and Particle Astrophysics of NTU and the National Center for Theoretical Sciences of Taiwan for the support.
\end{acknowledgments}


\end{document}